# Synchronous Observation on the Spontaneous Transformation of Liquid Metal under Free Falling Microgravity Situation

Xi-Hui He[1,2*], Li-Cong Zheng[2*], Qing-Shen Wu[2*], Zhi-Zhang Chen[2*], Zhi-Qiang Guan[2*], Mian Liu[2*], Ying-Bao Yang[3*], Lei Sheng[2,4], Ze-Jun Yang[3], Jing Liu[1,4,5**]

1. Institute for Advanced Study on Liquid Metal, Yunnan University, Chenggong, Yunnan Province 650504, China

2. Yunnan Jing Chuang Liquid Metal R&D Center, Qujing, Yunnan Province 652399, China

3. Yunnan Ke Wei Liquid Metal R&D Center, Xuanwei, Yunnan Province 655400, China

4. Key Lab of Cryogenics, Technical Institute of Physics and Chemistry, Chinese Academy of Sciences, Beijing 100190, China

5. Department of Biomedical Engineering, School of Medicine, Tsinghua University, Beijing 100084, China

\* These authors contribute equally to this research.
\*\* E-mail: jliu@mail.ipc.ac.cn

**Abstract**

The unusually high surface tension of room temperature liquid metal is molding it as unique materials for diverse newly emerging applications. However, unlike its practices on earth, such metal fluid would display very different behaviors when working in space situation where gravity disappears and surface property dominates the major physics. So far, few direct evidences are available to understand and fully utilize such effect which would impede further exploration of liquid metal uses for space. Here to preliminarily probe into this intriguing issue, a low cost experimental strategy to simulate microgravity environment on earth was proposed through adopting bridges with high enough free falling distance as the test platform. Then using digital cameras amounted along x, y, z directions on the outside wall of the transparent container with liquid metal and allied solution inside, synchronous observations on the transient flow and transformational activities of liquid metal can be performed. Meanwhile, an unmanned aerial vehicle was adopted to record the whole free falling dynamics of the test capsule from the far end which can help justify subsequent experimental procedures. A series of typical fundamental phenomena were thus observed as: (a) A relatively large liquid metal object would spontaneously transform from its original planar pool state into a sphere and float in the container if initiating the free falling; (b) The liquid metal changes its three-dimensional shape due to dynamic microgravity strength due to free falling and rebound of the test capsule; and (c) A quick spatial transformation of liquid metal immersed in the solution can easily be induced via external electrical fields. The mechanisms of the surface tension driven self-actuation of the liquid metal in space were interpreted. Further, some major factors and strategies to affect the phenomena were discussed. All these works indicated that microgravity effect should be fully treated in developing future generation liquid metal space technologies like smart fluidic system, shape changeable soft robot, chip cooling, energy management, advanced manufacture and biomedical utilizations etc.

## 1. Introduction

The room temperature liquid metals such as gallium or its alloy own many outstanding virtues that were not fully known before. This offers tremendous opportunities for developing advanced technologies in a group of newly emerging areas such as soft machine [1,2], chip cooling [3], microfluidics [4], printed electronics [5], stretchable device [6,7], functional switch [8], 3D printing [9], and bio-medical technology [10] etc. Among them, the transformation of liquid metal between different morphologies is especially receiving huge attentions due to its both interesting fundamentals and potential roles in innovating future generation shape changeable robot. With unusually large surface tension, say around 500-700 mN/m which is nearly ten times that of water [11], the liquid metal was recently found to display many unconventional performances [1]. In many of these practices, surface tension, as inherent characteristic of surfaces and interfaces of a material, plays the most critical role. It decides the shape and other dynamic characters of the liquid metal when placed on a free surface, immersed in allied solution or an interface with other condensed phase [12].

On earth, due to gravity effect, liquid metal generally appears as a pool of planar fluid if a large amount of such liquids were placed in a container [11]. However, when subject to microgravity environment, where objects appear to be weightless, such behaviors may become very different. According to the former experiences on the water droplet test in the space, one can envision that the object made of liquid metal may also just flexibly float as a sphere in the space. This would lead to game changing design and making of new conceptual devices and systems for space exploration. In fact, investigation on the microgravity effect to the properties and applications of various metals has been a core issue owing to rapid progresses of latest space explorations over the past few decades. Typical researches along this category are generally like: phase change [13-15], metal foam structure [16-18], shape and geometry [19, 20], interfacial behavior [21], levitation methodology [22, 23], physical property measurements [24, 25], mutual immiscibility [26], diffusion issue [27-30], and liquid bridge [31] etc. However, nearly all these endeavors are focused on the high melting point metals such as tin, lead, copper and the like. So far, academic efforts on the room temperature liquid metal and its applications in space or simulated microgravity environment are few, let alone the unique fluidic and transformational behaviors of liquid metal immersed in the solution or air. This therefore left rather diverse fundamentals and practical issues to be answered. For example, a most basic question would come to that, if the gravity effect becomes weak or even disappear, what will happen to the liquid metal. Such issues are deemed to dominate future space liquid metal systems and related applications. Clearly, to clarify these mysteries, realizing a microgravity environment is the first step.

As is widely known, Bond number is often adopted as the major dimensionless parameter to characterize the microgravity effect. It refers to the relative ratio between surface tension and gravity of an object [32] and reads as $Bo = \Delta \rho g L^3 / \sigma$, where $\Delta \rho = \rho_{metal} - \rho_{fluid}$ is density difference between liquid metal and the surrounding fluid such as solution or ambient air, $\sigma$ is surface tension of liquid metal inside the fluid, g is gravity acceleration speed, and L is characteristic size of the liquid metal. In principle, when $Bo \ll 1$, the ratio of gravity effect over surface tension becomes weak. From this aspect, three typical approaches can be adopted to simulate microgravity [32]: (1)

Finding liquid metal which has extremely high surface tension however small density. (2) Using matching liquid with as close density as that of liquid metal. (3) Another way is just taking a small liquid droplet so as to watch its microgravity effect such as that gallium droplet often appears as a sphere in NaOH solution. However, all these strategies are indirect ways. Clearly, the most straightforward way to investigate the microgravity issue is to create such an environment even for only a short period of time. Along this direction, typical experimental approaches are like [32]: space shuttle, parabolic flights, rocket, and drop tower etc. Without any doubt, all these strategies are expensive and could not be easily affordable by an ordinary lab.

Aiming to partially address the above tough issue and to establish an easy going pervasive way to probe into the potential activities of liquid metal in space, a low cost experimental method based on the free falling experimental principle was established. Such platform can be administrated as desired for the direct observation of the microgravity effect to the flow and transformational behaviors of liquid metal. As everybody knows, due to earth's gravity, any test platform would be pulled downward toward the surface with an ever increasing speed until hit the ground. The larger distance, the longer time for the test platform will sustain the microgravity state. For this reason, we have identified an idealistic natural condition through adopting bridge as the free falling test platform. In this study, two bridges were simultaneously introduced for the experiments. Particularly, the Bei Pan Jiang Bridge, situated in Xuanwei City, Yunnan Province, China, well known for its magnificent perpendicular falling distance of 565m, as high as 200 floor building, was adopted. This bridge, standing as the highest from its road surface to the bottom over the world, offers us an idealistic test zone to realize a relatively long enough observation on the microgravity effect of liquid metal transformation during free falling test. Meanwhile, another bridge with much shorter free falling distance as 20-30m was also fully used which provides rather convenient platform for performing many easy going tests as disired.

**2. Materials and methods**

Without losing any generality, in current experiments, we adopted the most commonly known eutectic liquid metal GaInSn alloy whose composition is 67% Ga, 20.5% In, and 12.5% Sn by volume as the test subject. Such alloy's melting point is 10.35 ℃ which is beneficial to always maintain in liquid state at around room temperature. The alloy was prepared from gallium, indium and tin metals with purity of 99.99 percent. For the preparation, the raw materials for the three metal components with a volume ratio of 67: 20.5: 12.5 were added into the beaker and heated at 100 ℃. After these metals were all melted, a magnetic stirrer was utilized to stir their mixture until homogeneous alloy was achieved.

As is well known, the Ga-based liquid metal is easy to be oxidized even in a trace amount of oxygen by forming $Ga_2O_3$ oxides on its surface (~1 nm in dry air). Such oxide film would evidently change the surface tension and then restrict the transformation of the liquid metal. To solve such an issue, the prepared liquid metals can be immersed into NaOH solution (0.5mol/L) which is to remove its oxides as could as possible. Nevertheless, if one wishes to regulate the surface tension, such liquid metal can also be partially exposed to the ambient air. The liquid metal and the solution were placed in a transparent container so that surrounding cameras can shoot the videos and fluidic events happening inside. According to the specific need, the container can be designed as different sizes and structures. Here two typical transparent containers used in current article are cube 10cm*10cm*10cm (Fig. 1(a)) and cuboid 10cm*10cm*20cm respectively. The container then would be amounted on a capsule for the subsequent free falling tests.

A representative experimental set up and test procedures can be found in Fig. 1. There, Fig. 1(a) schematically depicts the principle of the current free falling experiment. Fig. 1(b) gives different views on one of the two bridges, the Bei Pan Jiang Bridge as the test platform. Presented in Fig. 1(c), (d) are several test capsules ready for the free falling experiment. The whole system was assembled as shown in Fig. 1 (e). To completely shoot and record the transformation of the liquid metal object, three digitized cameras (SHETU 4K F68) with quick enough CCD speed were adopted to synchronously grasp the dynamic behaviors of the liquid metal and the surrounding fluid (Fig. 1(c)). Such imaging systems were fixed along x, y and z axial direction respectively in the outside walls of the capsule via an appropriate distance so as to guarantee that they can shoot the images of the tested subjects with as accurate as possible shape. Among them, two cameras were amounted at the perpendicular two sides of the capsule for grasping the videos along x and y directions and the third one was fixed on the top side wall of the transparent container to obtain the top view z directions video. Fig. 1(e) illustrates the moment for conducting the subsequent free falling test. Because the free falling experiment may endure a pretty large distance say about 400 m depth, an unmanned aerial vehicle was also introduced to shoot and record the whole dynamic process. This would help justify the falling state of the test capsule and thus prepare well for the subsequent experiments.

One thing should be of cautions in the designing and manufacture of the current test object is that its weight center should be placed to the lowest part as could as possible so as to guarantee the quality of the microgravity environment. Since the current system is designed for repeatable and multiple tests, we adopted a linked rope to drag the capsule which can be up to even 400 meters long. Before test, such rope should be dropped down in advance so as to avoid adding gravity on the test capsule. A most important core issue for the whole test is the safety issue. As can be anticipated, at the end of the free falling, a huge force will be produced which could easily break up the rope pulling up the capsule. For example, according to the law of momentum conservation, one can evaluate such force as $F_{max} t = mv_2 - mv_1$, where F is the ending force, t the duration time of the rope when subject to immediate stress, m the mass of the capsule, $v_1$, $v_2$ are respectively the stop and final velocity. If treating $v_1$ as zero, and considering current test capsule with weight around 30 kg and a falling distance h=400 m, the ending striking force can be even up to 5000 ton within the case of t~0.1-0.01s. This may possibly break up the rope and damage the whole test system. To avoid such thing to happen, we need to reduce the tremendous stress caused by the action at the end of the free falling. For this purpose, an elastic wire was particularly adopted as the ending wire which significantly buffers the striking force.

To test various specific cases when microgravity would take into effect, a series of designed experiments were performed and typical cases were listed in Table 1.

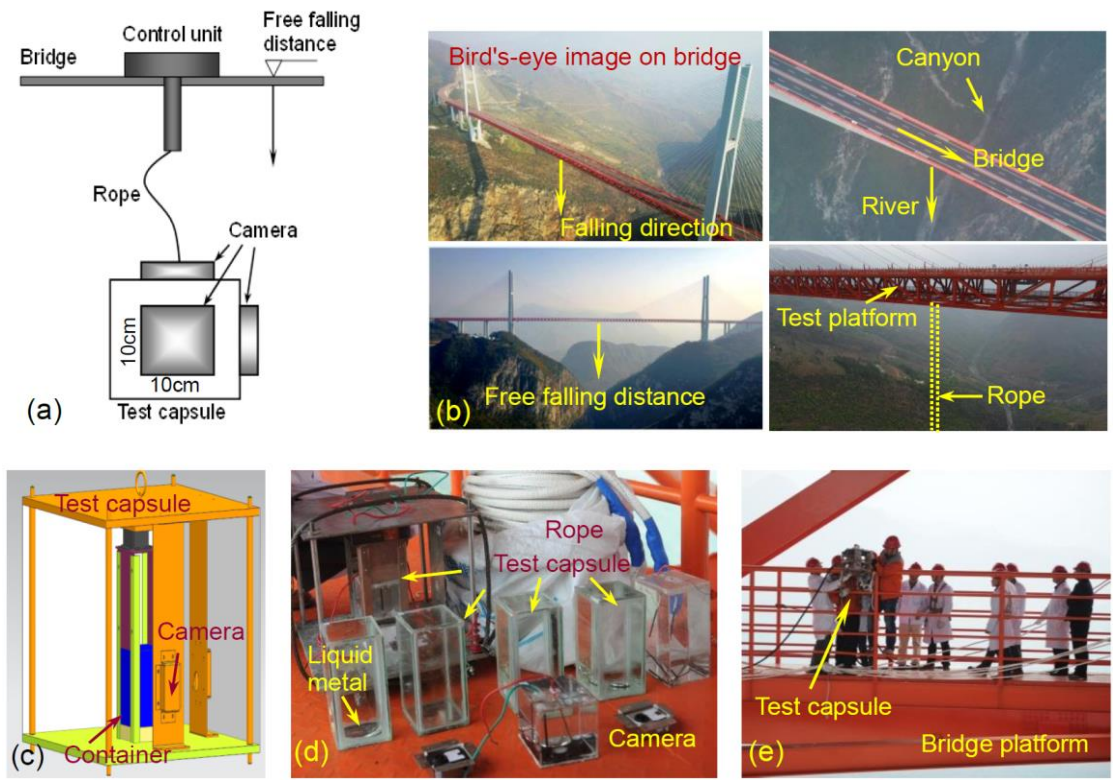

**Fig. 1** Schematic for the experimental set up for the synchronous observation on the microgravity effect to the spontaneous transformations of liquid metal via free falling experiments performed on a bridge with operating height up to 400 m. (a) Principle of free falling test; (b) Experimental platform set on the Bei Pan Jiang Bridge, Xuan Wei, China, whose largest height is 565 m; (c) Test capsule design; (d) A group of typical capsules for the tests; (e) Preparing free falling test on the bridge. (Note: The images were taken by either unmanned aerial vehicle or camera)

**Table 1** Typical cases for the liquid metal free falling experiments

| Group | Sample | Sample quantity | Electrical field |
|---|---|---|---|
| 1 | Air + Liquid metal | Small amount to form droplet | Off or on |
| 2 | Air + Liquid metal | Large droplet covering 1/10~2/3 container bottom | Off or on |
| 3 | Air + Liquid metal | Liquid metal covering container bottom with depth of about 0.5~1cm | Off or on |
| 4 | Liquid metal with NaOH on surface | Liquid metal covering container bottom with depth of about 0.5~1cm | Off or on |
| 5 | NaOH (Full) + Liquid metal | Small amount to form droplet | Off or on |
| 6 | NaOH (Full) + Liquid metal | Large droplet covering 1/10~2/3 container bottom | Off or on |
| 7 | NaOH (Full) + Liquid metal | Liquid metal covering container bottom with depth of about 0.5~1cm | Off or on |

## 3. Results

Presented in Fig. 2 are four typical snapshots of the transformation of small liquid metal droplet immersed in 0.5 mol/L NaOH solution contained in a transparent cube container with size of 10cm*10cm*10cm. At the initial state, the liquid metal just stands as a small planar pool (a). If initiating the free falling, a spheric liquid metal droplet would spontaneously form with the loss of gravity ((b), (c)). At the end of the free falling, the liquid metal droplet dances and get pressed like an elastic ball (d) due to strong striking on the bottom of the container until finally broke up into desperate smaller droplets (e), (f).

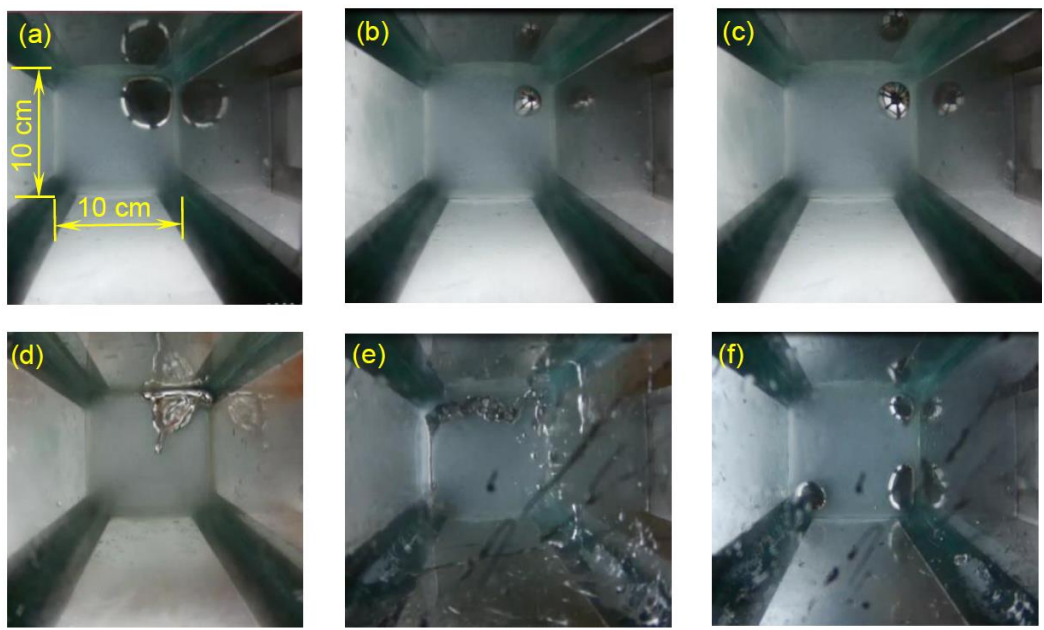

**Fig. 2** Typical snapshots of transformation of small liquid metal droplet immersed in NaOH solution in short term microgravity environment for about 2 s (Top view). (a) Initial state. (b), (c) A spheric liquid metal droplet forms with loss of gravity. (d) The liquid metal droplet dances and gets pressed due to strong striking on the bottom of the container. (e) The liquid metal was broken into desperate smaller droplets due to striking by the surrounding solution and container wall. (f) Ending state. (Test condition: Container size: 10cm*10cm*10cm; Bridge height: 20m; Ambient temperature 16 ℃; Humidity: 59%; Wind force: Light breeze)

If a further larger liquid metal object was tested, more vivid and profound transformation and fluidic behaviors of liquid metal can be observed. Fig. 3 depict several typical snapshots for the subsequent transformations of liquid metal pool immersed in NaOH solution over the transient free falling lasting for about 22s. Similar to the case of Fig. 2, at the starting time: 11: 09: 37, liquid metal appears as a planar pool covering the container bottom due to gravity. However, with the weakening of the gravity, the role of the surface tension becomes dominant and liquid metal automatically turns to stand up (b) and gradually forms a spheric liquid metal droplet (c), (d). Since it is impossible to maintain the capsule at the same configuration during the free falling process, the gravity should exist all the time, although strong or weak. This may therefore induce flows of the solution as well as the liquid metal. As a result, one can observe that the spheric liquid metal may contact with the transparent wall (e). Then, liquid metal returns back to its pool shape due to gravity and keeps oscillating (g)~(i) and goes up and down (j) if the rope was strongly bounced back at the end of the falling.

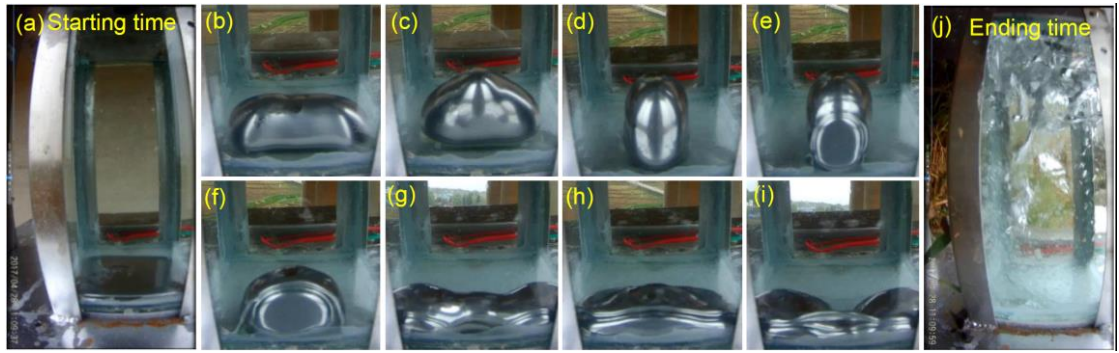

**Fig. 3** Typical snapshots for the spontaneous transformations of liquid metal pool immersed in NaOH solution over the transient free falling (Side view). (a) Starting time: 11: 09: 37 for liquid metal pool. (b) Liquid metal turns to go up when dropping off the test capsule; (c), (d) A spheric liquid metal droplet forms with loss of gravity. (e) The spheric liquid metal contacts with the container wall. (g)~(i) Liquid metal returns back to its original pool shape due to gravity and keeps oscillating. (j) Ending time 11: 09: 59, violent flow occurs inside the container with the test capsule going up and down. (Container size: 10cm*10cm*20cm)

As has been mentioned above, the solution may also add force to the transformation and flow of the liquid metal. Theoretically, if the liquid metal was just left in the air, potential force of the surrounding solution can be significantly reduced and the microgravity effect would become much stronger. Again, this prediction was demonstrated by current study. In the test of Fig. 4, only a thin layer of NaOH solution was covered on the liquid metal to avoid oxygenization so as to maintain its high surface tension. That means the test subject was placed in the air. Fig. 4 presents a group of transformations of a liquid metal pool in air over the short term free falling lasting for about 2s. And again, at the starting time (a), liquid metal appears as a small pool due to gravity. Then, an evident spheric liquid metal droplet forms and floats with the loss of gravity and transforms its shape due to dynamic change of the microgravity (b)~(g). The liquid metal droplet breaks up ((h),(i)) due to striking on bottom of the container when gravity was recovered, and then it was seriously stricken by the container bottom (f).

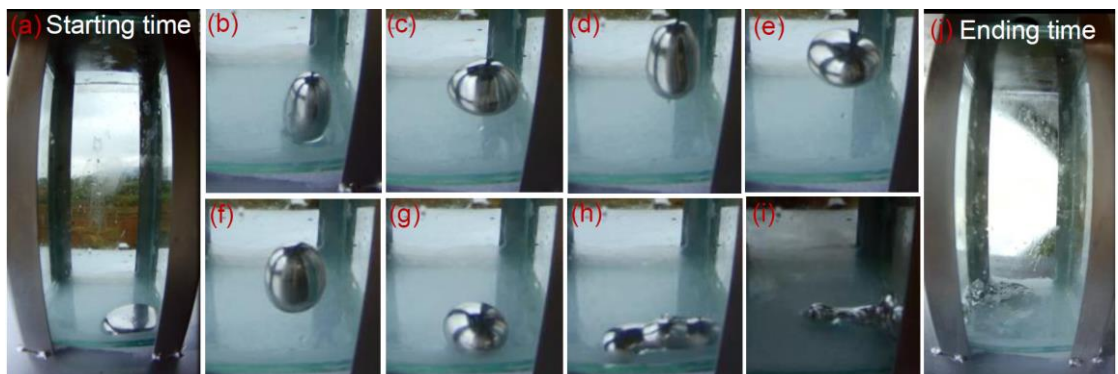

**Fig. 4** Typical snapshots for subsequent transformations of small liquid metal pool in the air over the short term free falling lasting for about 2s (Side view). (a) Starting time for liquid metal pool. (b)~(g) A spheric liquid metal droplet forms and floats during free falling. (h),(i) The liquid metal droplet breaks up. (j) Striking state of the liquid metal. (Container size: 10cm*10cm*20cm)

As has been revealed before [1], among all the external factors, electricity would easily induce the transformation of the liquid metal. It was found that under microgravity, such effect would become much stronger. As clarified by the tests presented in Fig. 5, typical snapshots indicate the transformations of liquid metal under electrical voltage 8.4V during the free falling. Unlike the transformation on earth, the electrical control now becomes rather agile and quick. For example, during the free falling, with loss of gravity, one can find that the electrode would immediately attract the liquid metal and a relatively large transformation will be caused ((b)~(e)). However, at the initial state on the ground, the same electrode with the same voltage could not even affect the same liquid metal at all. At the start time 12: 13: 18 (a) or ending time (f), the liquid metal just appears as a pool even the electricity was switched on (a). This finding reminds important value for space application in that, the electricity can be adopted to realize various easily controllable liquid metal transformers, machines or even integrated soft robot in the near future.

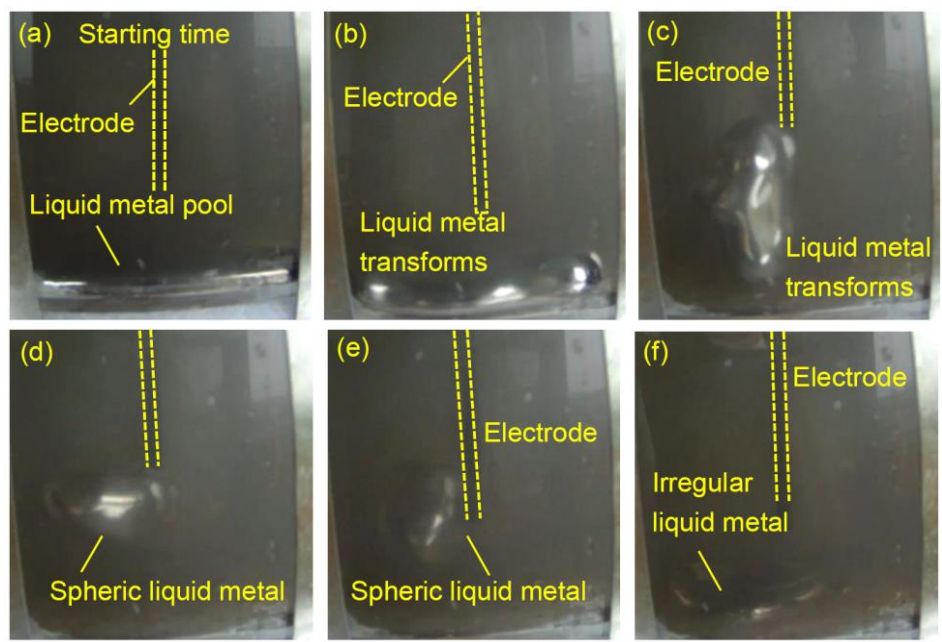

**Fig. 5** Typical snapshots for the transformations of liquid metal under electricity in free falling experiment (Side view). (a) Start time 12: 13: 18 for initial state of a pool of liquid metal. (b) Liquid metal turns to change shape with loss of gravity. (c) The liquid metal droplet was attracted to the electrode. (d),(e) The liquid metal subsequently transforms under electricity over the transient gravity effect. (f) Ending time: 12: 13: 18. (Container size: 10cm*10cm*10cm)

The above experiments were all carried on a not so long enough microgravity environment since the lasting time is only about 2s. Clearly, a relatively long term observation of the microgravity effect on the liquid metal transformation is extremely important. For this purpose, we turn to perform additional investigation using free falling experiments with significantly expended spanning time. The tests were conducted on Bei Pan Jiang Bridge whose largest height is 565m. And profound experimental results were obtained. Fig. 6 showed several typical snapshots of the transformations of liquid metal pool immersed in NaOH over the 34 s test. At the initial state, the liquid metal stays as a pool (a). During the falling, the liquid metal would produce discrete spheric liquid metal droplets with loss of gravity (b)~(f). Then transformation of liquid metal occurs due to imperfect free falling of the test capsule. In this case, partial microgravity and partial solution induced force will both play roles in the transformation of liquid metal. Therefore one can observe that, some of the liquid metal just breaks up into separate droplet due to violent action from the surrounding solution during the free falling (d) while some just appears as a pool. Transformation of the liquid metal happens due to reverse jumping of the test capsule. Whenever microgravity appears, separate spheric liquid metal droplets would form due to rebound of the test capsule (f), (g). Finally, the liquid metal returns to pool state.

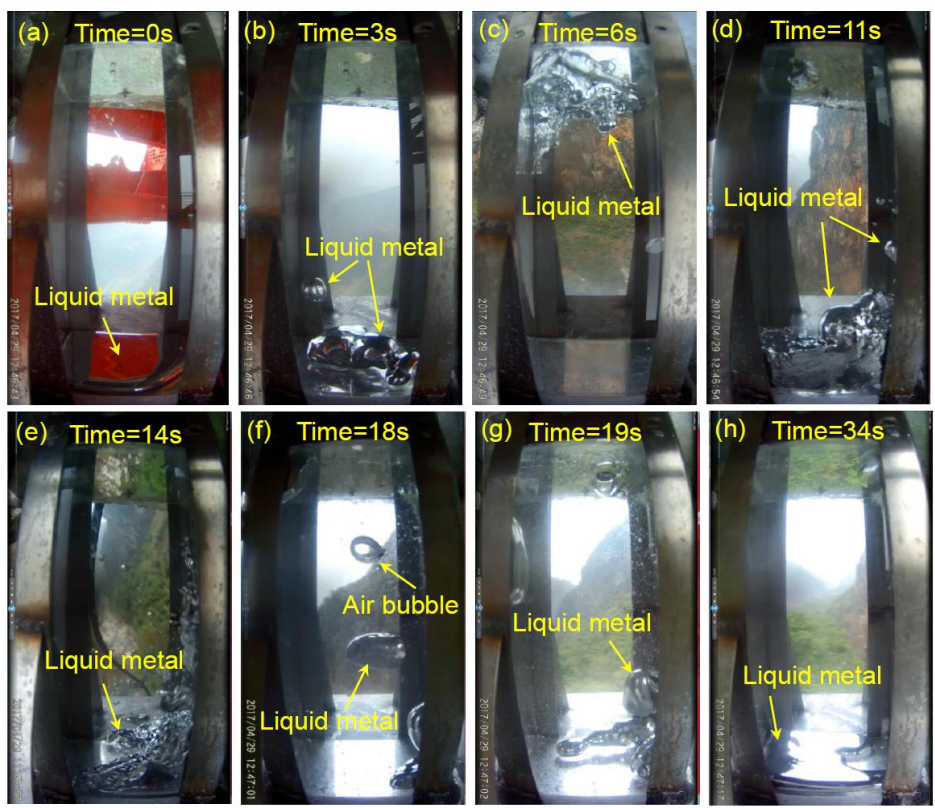

**Fig. 6** Typical snapshots of transformation of liquid metal pool immersed in NaOH in a relatively long term free falling environment (Side view). (a) Initial state. (b) Discrete spheric liquid metal droplets form with loss of gravity. (c) Transformation of liquid metal. (d) Liquid metal breaks up into separate droplet like objects due to ending of the free falling. (e) Transformation of liquid metal due to reverse jumping of test capsule. (f), (g) Separate spheric liquid metal droplets form with rebound of the test capsule. (h) Ending state. (Test condition: Container size: 10cm*10cm*20cm; Free falling height: 360m; Ambient temperature 14 ℃; Humidity: 56%; Wind force: Light breeze)

If administrating electricity on the test object, the long term observation on the liquid metal transformation can provide more dynamic events. Fig. 7 presents typical snapshots of transformation of liquid metal pool immersed in NaOH under electrical voltage 8.4V. The liquid metal would change its initial pool state (a) into separate droplets due to microgravity effect and electrical actuation (b), (c). Discrete spheric liquid metal droplets form with the loss of gravity. The liquid metal breaks up into separate droplet like objects due to ending of the free falling (d) and transform due to reverse jumping of the test capsule (e). Over the subsequent times, separate spheric liquid metal droplets form with rebound of the test capsule (g). At the ending steady state of the test capsule, the liquid metal just get back to its initial pool state (h), which is similar as that in initial state (a).

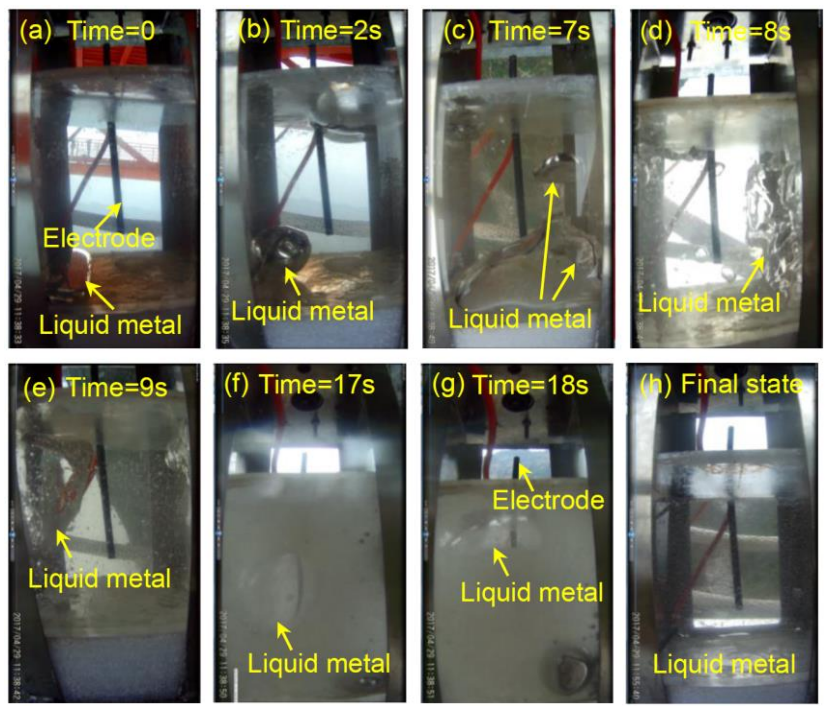

**Fig. 7** Typical snapshots of transformation of liquid metal pool immersed in NaOH under electricity in a relatively long free falling experiment (Side view). (a) Initial state. (b), (c) Separate liquid metal droplet forms due to microgravity effect and electricity. (d), (e) Transformation of liquid metal due to reverse jumping of the test capsule. (f), (g) Separate spheric liquid metal droplets form with rebound of the test capsule. (h) Ending steady state. (Test condition: Container size: 10cm*10cm*20cm; Free falling height: 300m; Ambient temperature 14 ℃; Humidity: 56%; Wind force: Light breeze)

## 4. Discussion

Under microgravity in space environment, the surface tension plays perhaps the most important role in dominating various surface and interface phenomena. As is well known, water has a higher surface tension (72.8 mN/m at 20 ℃) compared to most of other liquids. But this value is still much smaller than that of the liquid metal. It is also partially due to this reason, the room temperature liquid metal display rather unique properties and increasingly realized merits. As is well known, the surface tension of a fluid tends to make it acquire the least surface area. In microgravity environment, surface tension plays the core role in dominating the shape and transformation of the liquid while the mass of such object does not matter evidently. Therefore, the pool of the liquid metal can be found to spontaneously transform into a sphere when free falling happens. In addition, since the surface tension of liquid metal is much larger than that of water, the liquid metal pool is much easier to become a sphere while the surrounding NaOH solution does not.

For a free falling experiment, the duration time for maintaining the microgravity environment depends on the height of the falling distance which can be calculated as $\tau_\mu=(2h/g_e)^{1/2}$, where h is the falling height, $g_e$ is gravity acceleration rate. Such experiments are generally conducted in a vacuum tower in order to avoid air resistance. But for current liquid metal cases whose surface tension is orders larger than that of water, a slightly reduced gravity will clearly distinguish the effect of the strong surface tension to the basic transformational behaviors of the liquid metal. Therefore it can be found that, both tests on bridges either short or high successfully offer the necessary microgravity conditions.

In this study, all the transformations of the liquid metal are resulted from the breakup of the equilibrium between its surface tension and fluidic properties. For the liquid metal at initial steady state, its pressure gradient equals to gravity [32], i.e. $\triangle p=\rho g$. With the loss of the gravity, this equilibrium state will subject to spontaneous change and the solution as well as liquid metal will be induced to transform and flow. Regarding the pressure difference across the spheric liquid metal surface, one can write out its control equation as $\triangle p=p_i-p_e=2\sigma/r$, where $p_i$ and $p_e$ are internal and external pressures of the droplet respectively, and r is the radius of the droplet. During the free falling, the external pressure $p_e$ is in fact changeable. Therefore, the liquid metal will not always maintain in spheric shape. That is the reason why very different liquid metal profiles were observed. And if the external pressure is strong enough such as striking occurs at the end of the free falling, the liquid metal will be broken into separate droplets. It should be pointed out that, the whole process in the present experiments does not maintain very stable. This is caused due to disturbance during the acceleration and vibration of the test platform.

In most cases of the present study, the transparent container usually was completely filled with solution. Without influence from the induced solution flow, the liquid metal tends to easily form a sphere. However, in reality, there always exists certain air gap inside the container. Under microgravity, such bubble may float into the solution and add additional force on the liquid metal which will balance the microgravity effect. That is why liquid metal display rather irregular shape although under free falling. Such situation can in fact be disclosed through comparative conceptual experiments. If there is large air bubble inside the test capsule, it would add force on the liquid metal immersed inside the solution. In this case, the liquid metal may not easily form a sphere although a microgravity environment has already been established. This well explains that the long distance free falling experiment sometimes may not be able to offer more chances to watch the surface tension effect to the transformation of the liquid metal. Fig. 8 presents such evidences via comparative conceptual experiments. There, at initial state (a), we intentionally leave a large air space inside the test capsule with other spaces filled with NaOH solution. It can be found that, with the loss or weakening of the gravity, the air gap would be formed into a large sphere and generate forces on the supporting solution, which may not be uniform over the whole domain of the container ((b)-(e)). As a result, the liquid metal immersed inside may subject to forces of the surrounding solution from multiple directions. And it will not be easy for the liquid metal to transform into a sphere like object. As an alternative, one could completely remove the air bubble inside the test capsule (f). In this case, the solution inside tends to maintain equilibrium force and the liquid metal will almost take the similar shape as that in the vacuum.

The value as reminded by the present free falling experimental strategy is rather evident. Although gravity does become weaker when an object goes far enough from the earth, a small amount of gravity should always exist everywhere in the space. That means, zero gravity in fact does not exist. In this sense, the present results are of direct reference for the coming practices of liquid metal utilizations in a group of future space exploration, either near or far from the earth.

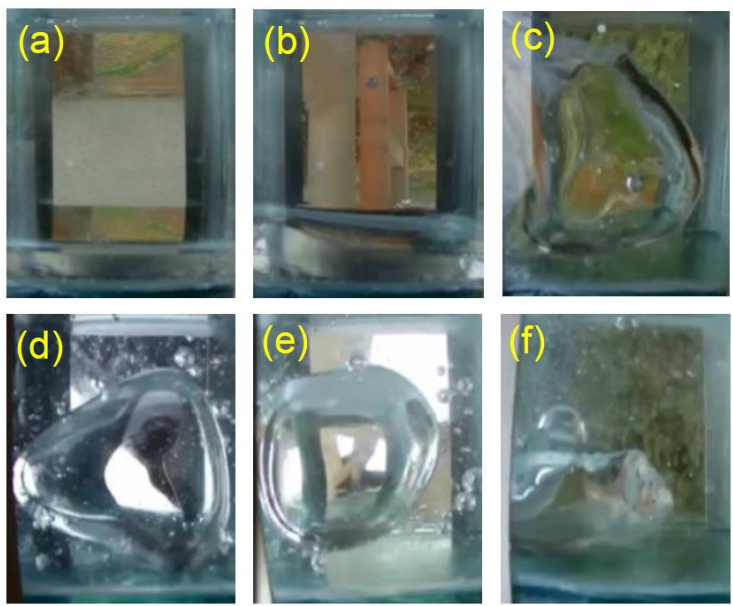

**Fig. 8** Typical snapshots of transformations of air bubble in short term microgravity environment lasting for 2s (Side view). (a) Initial state as rectangular air gap. (b) Air gap became immersed into the solution with loss of gravity. (c)~(e) A large spheric air bubble was formed inside the solution under microgravity. (f) Air bubble gradually returns back to its original state. (Container size: 10cm*10cm*20cm)

However, it should be pointed out that, the present study is only a preliminary investigation towards understanding the microgravity effect on the transformation of liquid metal. Many mechanisms especially quantification of the activities due to gradually changing microgravity, electrical field, solution effect, experimental apparatus, test climate conditions etc. remain unclear. A series of scientific and practical issues are waiting for clarification in the coming time. As the is initial endeavor to probe into the categories lying behind, the present experimental method provides a low cost way for quickly investigating physics, fluids and material issues of liquid metal under microgravity as one desired. In fact along this direction, many different natural environments such as a building, a mountain, or an unmanned aerial vehicle, etc. can all be adopted as the way to generate free falling microgravity effect. And, a group of experimental conditions can be improved. For example, the video shooting speed of the current camera system is still not quick enough, which omits many important details of the quick transformation and transition of the liquid metal. As an alternative, fast CCD camera is necessary and should be adopted in the coming study to record the quickly changing event of the liquid metal. Further, effects of wire and orientation to the microgravity situation are also very critical. It is known that, gravity center of the test capsule dominates the lasting time of the microgravity environment. With the variation of such center, gravity of the solution may add force on the liquid metal. As a result, surface tension effect of the liquid metal will not be displayed.

The liquid metal driving mechanism is related to the imbalance between the surface tension and the reduced gravity. This finding reminds possible ways for developing liquid metal soft machine, surface tension-driven motors, biomedical application and fluidic systems etc. under microgravity environment. On earth, the strong gravity force would significantly impede the flow and transformation of the liquid metal. Without such effect, one can manipulate more flexibly the liquid metal via a rather convenient way which would significantly expend the application of the liquid metal technologies. At this stage, many potential ways exist to regulate the surface tension which would enable space application of the liquid metal. For example, synthetically chemical-electrical mechanism (SCHEME) has been proposed to control large scale reversible deformation of liquid metal objects [10]. Such reversible configurational transformation is achieved via the combination of electrochemical oxidation and chemical dissolution processes of oxide gallium, and upon application of the external electric field and alkaline or acidic electrolyte solution, the surface tension of the liquid metal can be easily tuned within a wide range from 700mN/m to near zero. This allows people to develop more shape changing liquid metal machine, fluidic system and even robot (Fig. 9). Further, additional external factors such as magnetic field [33], ionic concentration [34] etc., as have been found to effectively affect the propelling behavior of liquid metal, are also worth of trying to intentionally alter the surface tension actuation capabilities in space. Besides, the self-powered liquid metal machine as disclosed in [35] would also provide profound opportunities for potential uses in developing future smart space soft machines or hydrogen energy generation system. All these issues can be investigated under microgravity by way of the current experimental approach.

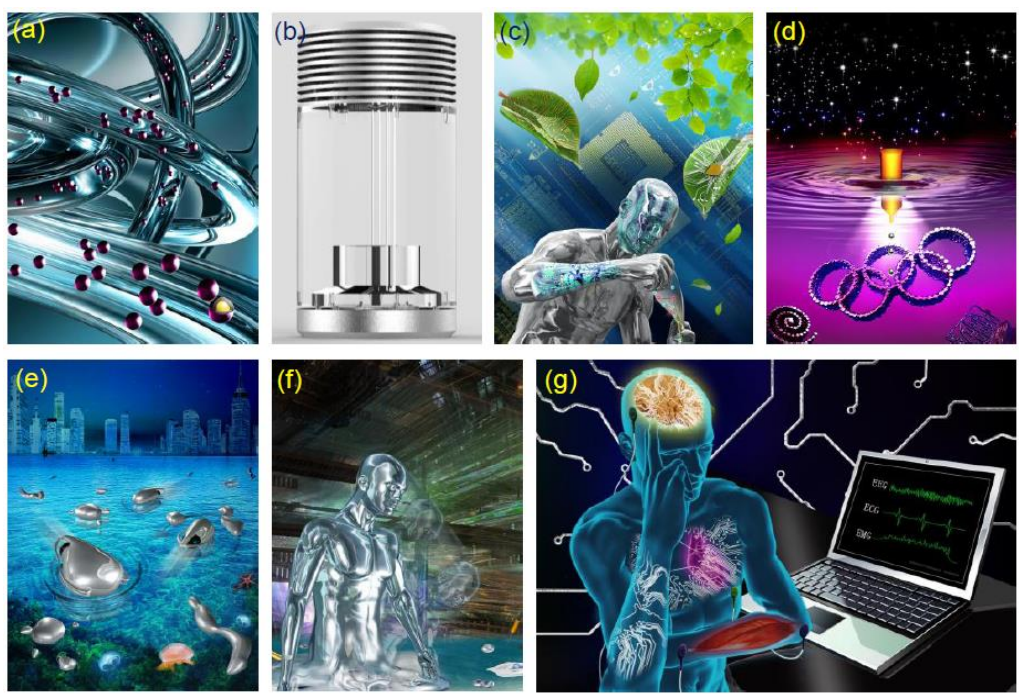

**Fig. 9** Important issues that microgravity might affect for the use of liquid metal in developing future generation space technologies. (a) Fluidics; (b) Heat transport; (c) Printed electronics; (d) 3D printing; (e) Soft machine; (f) Transformational soft robot; (g) Biomedical utilizations.

Overall, the surface properties of the liquid metal are rather vital in nearly all applications involved from chip cooling, thermal energy harvesting, hydrogen generation, shape changeable soft machines, printed electronics to 3D fabrication, etc. owing to its pretty large surface tension. Clearly, its further study and application for the space environment should be a core issue in the coming time. Potential application of liquid metal is huge (Fig. 9). Especially, a group of soft machine can be designed and tested for future use in space such as in The International Space Station. In the vacuum microgravity, any heavy objects can float and move around easily in the space. This warrants many potential uses of liquid metal. In certain extreme situation such as inside biological blood vessel system, it is hard to move the liquid metal via electricity. However, without the gravity, electricity can possibly be adopted to conveniently control the flow and transformation inside the vessel. The soft machine with more complicated performances in space can thus be made. A core mechanism lies in that the surface tension tends to mold liquid metal into a spheric droplet. Then one can ask such a basic question, i.e. since the liquid metal has a huge surface tension, under what circumstances, will the liquid metal always stands as sphere, whenever its size is small or large. All these intriguing situations and applications request further clarifications in the coming microgravity tests.

## 5. Conclusion

In summary, a low cost way for the synchronous observation on the transformations of liquid metal under simulated microgravity situation has been demonstrated on earth through introducing the free falling experiment on a bridge. A long enough duration of low gravity has thus been easily obtained which offers valuable moment to directly watch the spontaneous transformations of liquid metal caused by its strong surface tension effect. A series of comparative tests regarding short term or relatively long term microgravity situation performed on two bridges with very different free falling height disclosed important flow and transformational phenomena in the specific situations. Further, effects of the solution and external electrodes to the liquid metal transformations were also investigated. Overall, a surprising fact as revealed by this research lies in that, during the microgravity environment, the large pool of liquid metal would easily transform into a perfect sphere, although such cases only happen for small liquid metal droplet on earth. And electricity is rather efficient in controlling the transformation of liquid metal in microgravity. Starting from this point, one can envision that the liquid metal would play many potential roles for future space exploration. The present experiments also offer knowledges that are hard or even impossible to envision via an ordinary ground test which could help get the basic knowledge of liquid metal especially surface tension dominated behaviors in space environment. Finally, some potential applications for the coming space endeavor were suggested. Important fundamental and practical issues as raised by this study were also proposed. The present work is expected to serve as an important reference for the coming intense study regarding microgravity effects on liquid metal via a straightforward and efficient way on earth.

**Acknowledgements**

The authors gratefully appreciate the valuable help on the experiments from colleagues at Jing Chuang Liquid Metal R&D Center, Ke Wei Liquid Metal R&D Center, as well as the support regarding safeguard and permission for the tests on Bei Pan Jiang Bridge, including unmanned aerial vehicle and camera recording from colleagues at Xuan Wei City. This work is partially supported by Dean's Research Funding and the Frontier Project of the Chinese Academy of Sciences.